\documentclass[mathleft
]{an}
\usepackage{graphicx}
\usepackage{times}

\begin{document}

\Pagespan{789}{}
\Yearpublication{0000}%
\Yearsubmission{00005}%
\Month{00}%
\Volume{000}%
\Issue{00}%

\title{Re-determining the Galactic spiral density wave\\
       parameters from data on masers with \\
       trigonometric parallaxes}

\author{A.T. Bajkova\inst{1}\fnmsep\thanks{Corresponding author:
  \email{anisabajkova@rambler.ru}\newline}
\and  V.V. Bobylev\inst{1,2} }
\titlerunning{Re-determining the Galactic spiral density wave
       parameters}
\authorrunning{A.T. Bajkova \& V.V. Bobylev}
\institute{Central (Pulkovo) Astronomical Observatory, Pulkovskoye
Shosse 65/1, St.-Petersburg, 196140, Russia \and Sobolev
Astronomical Institute, St.-Petersburg State University,
Universitetskii pr. 28, Petrodvorets, 198504, Russia}

\received{00 Mon 0000} \accepted{00 Mon 0000} \publonline{later}


\abstract{%
The parameters of the Galactic spiral wave are re-determined using
a modified periodogram (spectral) analysis of the galactocentric
radial velocities of 58 masers with known trigonometric
parallaxes, proper motions, and line-of-site velocities. The
masers span  a wide range of galactocentric distances,
$3<$R$<14$~kpc, which, combined with a large scatter of position
angles $\theta$ of these objects in the Galactic plane $XY$,
required an accurate account of logarithmic dependence of
spiral-wave perturbations on both galactocentric distance and
position angle. A periodic signal was detected corresponding to
the spiral density wave with the wavelength $\lambda=2.4 \pm 0.4$
kpc, peak velocity of wave perturbations $f_R=7.5 \pm 1.5$ km
s$^{-1}$, the phase of the Sun in the density wave
$\chi_\odot=-160 \pm 15^\circ$, and the pitch angle of $-5.5 \pm
1^\circ$. }

\maketitle

\section{Introduction}

Spectral analysis of residual velocities of different young
galactic objects (HI clouds, OB stars, open star cluster younger
than 50 Myr, masers) tracing the Galaxy spiral arms has been
fulfilled, for example, by Clemens (1985), Bobylev, Bajkova \&
Stepanishchev (2008), Bobylev \& Bajkova (2010). As a result there
were determined the following parameters of the spiral density
wave subject to the theory by Lin \& Shu (1964): amplitude and
wavelength of the perturbations, evoked by the spiral wave, pitch
angle, phase of the Sun in the spiral wave. Nowadays galactic
masers having high-precision trigonometric parallaxes,
line-of-sight velocities and proper motions (Reid et al.~2009,
Rygl et al.~2010) are of great interest.

The previous spectral analysis represents the simplest periodogram
analysis of velocity perturbations based on conventional Fourier
transform, what can be considered only as the first approximation
of  exact spectral analysis and can be applied adequately only in
the case of small range of galactocentric distances (2-3 kpc). But
analysis of modern data on galactic masers which are located in
wide range of galactocentric distances ($3<$R$<14$~kpc) requires
elaboration of more correct tools of spectral analysis accounting
both logarithmic dependence from galactocentric distances and
position angles of objects. For a detailed description of the new
method of spectral analysis of velocity residuals see Bajkova \&
Bobylev (2012).

Our first study (Bobylev \& Bajkova 2010) was based on an analysis
of radial galactocentric velocities of only 28 Galactic masers.
The second one (Bajkova \& Bobylev 2012) dealt with 44 masers.
Currently, high-precision VLBI measurements of parallaxes,
line-of-site velocities, and proper motions are available for 58
Galactic masers, which is of great interest for our task. The aim
of this present study is to re-determine the spiral density wave
parameters by applying the recently proposed algorithm (Bajkova \$
Bobylev 2012) and new ones, described below, to more extensive
data series.

\section{Basic relations}

The velocity perturbations of Galactic objects produced by a
spiral density wave (Lin \& Shu 1964) are described by the
relations
\begin{equation}
\label{e-01}
V_R = - f_R \cos\chi,
\end{equation}
\begin{equation}
\label{e-02}
\Delta V_{\theta} = f_{\theta} \sin\chi,
\end{equation}
where
\begin{equation}
\label{e-03}
\chi = m[\cot(i)\ln(R/R_{\circ})-\theta]+\chi_{\odot}
\end{equation}
is the phase of the spiral density wave; $m$ is the number of
spiral arms; $i$ is the pitch angle; $\chi_{\odot}$ is the phase
of the Sun in the spiral density wave (Rohlfs 1977); $R_{\circ}$
is the galactocentric distance of the Sun; $\theta$ is the
object's position angle: $\tan\theta = y/(R_\circ-x)$, where $x,$
$y$ are the Galactic heliocentric rectangular coordinates of the
object; $f_R$ and $f_\theta$ are the amplitudes of the radial and
tangential perturbation components, respectively; $R$ is the
distance of the object from the Galactic rotation axis, which is
calculated using the heliocentric distance $r=1/\pi$:
\begin{equation}
\label{e-003}
R^2=r^2\cos^2 b-2R_\circ r\cos b\cos l+R^2_\circ,
\end{equation}
where $l$ and $b$ are the Galactic longitude and latitude of the
object, respectively.

Equation (\ref{e-03}) for the phase can be expressed in terms of
the perturbation wavelength $\lambda$, which is equal to the
distance between the neighboring spiral arms along the Galactic
radius vector. The following relation is valid:
\begin{equation}
\label{e-04}
\frac{2\pi R_{\circ}}{\lambda} = m\cot(i).
\end{equation}
Equation (\ref{e-03}) will then take the form
\begin{equation}
\label{e-05}
\chi = \frac {2\pi R_{\circ}}{\lambda}
\ln(R/R_{\circ})-m\theta+\chi_{\odot}.
\end{equation}

The question of determining the residual velocities is considered
below. To goal of our spectral analysis of the series of measured
velocities $V_{R_n}, \Delta V_{\theta_n}$ $n=1,2,\dots,N,$ where
$N$ is the number of objects, is to extract the periodicity in
accordance with model~(\ref{e-01})-(\ref{e-02}) describing a
spiral density wave with parameters $f_R,f_\theta,$ $\lambda,$ and
$\chi_\odot$. If the wavelength $\lambda$  is known, then the
pitch angle $i$ is easy to determine from Eq.~(\ref{e-04}) by
specifying the number of arms $m$. Here, we adopt a two-armed
model, i.e., $m=2$.

\section{Methods}

\subsection{Fourier transform-based analysis}

Let us represent series of velocity perturbations of galactic
objects evoked by a spiral density wave (\ref{e-01})-(\ref{e-02})
in the most general, complex form:
\begin{equation}
\label{e-06}
V_n=V_{R_n}+j\Delta V_{\theta_n},
\end{equation}
where $j=\sqrt-1$, $n$ is a number of an object ($n=1,...,N$).

A periodogram analysis, which we consider here, requires
calculation of power spectrum of series (\ref{e-06}) expanded over
orthogonal harmonic functions
$$ \exp[-j\frac{2\pi
R_{\circ}}{\lambda_k}\ln(R_n/R_{\circ})+jm\theta_n]
$$ in
accordance with expression (\ref{e-05}) for the phase.

A complex spectrum of our series is:
\begin{eqnarray}
\label{e-006}
\bar{V}_{\lambda_k}=\bar{V}_{\lambda_k}^{Re}+j\bar{V}_{\lambda_k}^{Im}
=\frac{1}{N}\sum_{n=1}^N (V_{R_n}+j\Delta
V_{\theta_n})\times\\
\times \exp[-j\frac{2\pi R_{\circ}}{\lambda_k}\ln(R_n/R_{\circ})+j
m\theta_n]\nonumber,
\end{eqnarray}
where the upper indices $Re$ and $Im$ designate real and imaginary
spectrum parts respectively.

Let us reduce the latter expression to a standard discrete Fourier
transform in the following way:
\begin{eqnarray}
\label{e-07}
\bar{V}_{\lambda_k}=\frac{1}{N}\sum_{n=1}^N
(V_{R_n}+j\Delta
V_{\theta_n})\times \exp(jm\theta_n)\times \\
\times \exp[-j\frac{2\pi
R_{\circ}}{\lambda_k}\ln(R_n/R_{\circ})]=\nonumber
\\
=\frac{1}{N}\sum_{n=1}^N V_n^{'}\exp[-j\frac{2\pi
R_{\circ}}{\lambda_k}\ln(R_n/R_{\circ})]\nonumber,
\end{eqnarray}
where

\begin{eqnarray}
\label{e-08}
V_n^{'}=V_n^{Re}+V_n^{Im}=\\=[V_{R_n}\cos(m
\theta_n)-\Delta
V_{\theta_n}\sin(m\theta_n)]+\nonumber\\
+j[V_{R_n}\sin(m \theta_n)+\Delta V_{\theta_n}\cos(m
\theta_n)]\nonumber.
\end{eqnarray}
And, finally, making the following change of variables
\begin{equation}
\label{e-09}
R_n^{'}=R_{\circ}\ln(R_n/R_{\circ}),
\end{equation}
we obtain standard Fourier transform of a new series $V_n^{'}$
(\ref{e-08}), determined in point set $R_n^{'}$:

\begin{equation}
\label{e-10}
\bar{V}_{\lambda_k}=\frac{1}{N}\sum_{n=1}^N
V_n^{'}\exp[-j\frac{2\pi R_n^{'}}{\lambda_k}].
\end{equation}

The periodogram $|\bar{V}_{\lambda_k}|^2$ is subject to further
analysis. The peak of the periodogram determines the sought-for
periodicity. The coordinate of the peak gives the wavelength
$\lambda$ and, respectively, pith angle $i$ (see Eq.(\ref{e-04})).
Relation between a peak value of the periodogram $S_{peak}$ and
perturbation amplitudes $f_R$ and $f_\theta$ is expressed as
follows:
\begin{equation}
\label{e-100}
 f_R^2+f_{\theta}^2=2\times S_{peak}.
\end{equation}

It is necessary to note that the spectral analysis of complex
series (\ref{e-06}) allows to determine $\lambda$ (or pitch angle
$i$) and phase of the Sun $\chi_{\odot}$, but does not allow to
estimate the amplitudes $f_R$ and $f_\theta$ separately. To
determine them it is necessary to analyze radial $\{V_{R_n}\}$ and
tangential $\{\Delta V_{\theta_n}\}$ velocity perturbations
independently, as it has been shown by Bajkova \& Bobylev (2012).
Here we show how amplitudes of perturbations can be found if
$\lambda$ and $\chi_{\odot}$ are known (for example, from previous
complex analysis). We consider series $\{V_{R_n}\}$ (by analogy
the same algorithm can be applied to series $\{\Delta
V_{\theta_n}\}$).

Let us represent Eq.~(\ref{e-03}) as
\begin{equation}
\label{e-11}
\chi = \chi_1-m\theta,\label{e-11}
\end{equation}
where
\begin{equation}
\label{e-12}
\chi_1 = \frac {2\pi
R_{\circ}}{\lambda}\ln(R/R_{\circ})+\chi_{\odot}.
\end{equation}
Substituting (\ref{e-11}) into Eq. (\ref{e-01}) for the
perturbations at the $n$th point and performing standard
trigonometric transformations, we will obtain
{\setlength{\mathindent}{0pt}
 \begin{eqnarray}
 \label{e-13}
      V_{R_n}=-f_R \cos(\chi_{1_n} - m\theta_n)=\\
      =-f_R\cos\chi_{1_n}\cos m\theta_n   - f_R\sin\chi_{1_n}\sin m\theta_n =\nonumber\\
      =-f_R \cos\chi_{1_n}(\cos m\theta_n + \tan\chi_{1_n}\sin
m\theta_n)\nonumber.
 \end{eqnarray}}

Let us designate
\begin{equation}
\label{e-14}
 V^{'}_{R}=-f_R\cos\chi_1,
\end{equation}
Owing to the substitution (\ref{e-14}), it then follows from
(\ref{e-13}) that
\begin{equation}
\label{e-15}
V_{R_n}=V^{'}_{R_n}(\cos m\theta_n+\tan\chi_{1_n}
\sin m\theta_n).
\end{equation}
Substituting known values of $\lambda$ and $\chi_{\odot}$ we can
form from Eq. (\ref{e-15}) a new data series
\begin{equation}
\label{e-16}
V^{'}_{R_n}=V_{R_n}/(\cos m\theta_n+\tan\chi_{1_n}
\sin m\theta_n).
\end{equation}
Again, using the substitution (\ref{e-09}), we obtain a standard
Fourier transform:

\begin{equation}
\label{e-17}
 \bar{V}_{\lambda_k} = \frac{1} {N}\sum_{n=1}^{N} V^{'}_{R^{'}_n}
 \exp\Bigl(-j\frac {2\pi R^{'}_n}{\lambda_k}\Bigr).
\end{equation}

In this case relation between a peak value of the periodogram
$S_{peak}$ and perturbation amplitudes $f_R$ is as follows:

\begin{equation}
\label{e-18} f_{R}^2=4\times S_{peak}.
\end{equation}

Note, that a separate periodogram analysis based on operations
(\ref{e-11})-(\ref{e-17}) can be realized as an iterative process
of seeking for unknowns $\lambda$, $\chi_{\odot}$, and $f_{R}$
under optimization of some specific signal extraction quality
criterium (Bajkova \& Bobylev, 2012).

For numerical realization of Fourier transform (\ref{e-10}) or
(\ref{e-17}) using fast Fourier transform (FFT) algorithms  it is
necessary to determine  data $V_{n}^{'}(R^{'}) (n=1,\dots,N)$ on
discrete grid $l=1,\dots,K=2^\alpha$, where $\alpha$~ is integer,
positive, $N\le K$; $\Delta_R$ is a discrete space. Coordinates of
data are determined as follows:
$l_n=[(R_n^{'}+|\min\{R_k^{'}\}|_{k=1,...,N})/\Delta_R]+1,
n=1,...,N$, where $[a]$ denotes an integer part of $a$. The
sequence determined is considered as a periodical one with the
period $D=K\times\Delta_R$. Obviously, the values of the
$K$--point sequence are taken to be zero in the pixels into which
no data fall.

\subsection{The GMEM-based analysis}

So far we have considered the simplest method of periodogram
analysis based on linear Fourier transform. In the case where the
data series are irregular, i.e., there are large gaps, the signal
spectrum is distorted by large side lobes and it becomes difficult
to distinguish the spectral component of the signal from spurious
peaks. In this case, it may turn out to be useful to apply
nonlinear methods of spectrum reconstruction from the available
data. This problem is fundamentally resolvable if the sought for
signal has a finite spectrum. Since our problem belongs to the
class of problems on the extraction of polyharmonic functions from
noise, we assume that this condition is met.

Here we propose a complex spectrum reconstruction algorithm based
on well-known maximum entropy method (MEM). Since the spectrum is
described by a complex-valued function, we apply a generalized
form of MEM (GMEM) proposed and described in detail by Bajkova
(1992) and Frieden \& Bajkova (1994).

The spectrum $\bar{V}_k=\bar{V}_k^{Re}+j\bar{V}_k^{Im}$ and the
data $V_{n}^{'}$ are related by the inverse Fourier transform:
{\setlength{\mathindent}{0pt}
\begin{eqnarray}
\label{e-19}
 \sum_{k=1}^{K} (\bar{V}_k^{Re}+j\bar{V}_k^{Im})
 \exp(j\frac {2\pi (k-1)(l_n-1)}{K})=V_{n}^{'}.
\end{eqnarray}}

Note that wavelength $\lambda_k$ and spatial frequency $(k-1)$ are
related as follows:
\begin{equation}
\label{e-190}
 \lambda_k=\frac{D}{k-1}.
\end{equation}

In our case, the reconstruction problem assumes finding the
minimum of the following generalized entropy functional:
{\setlength{\mathindent}{0pt}
\begin{eqnarray}
\label{e-20} E=\sum_{k=K_1}^{k=K_2} V_k^{Re+}\ln(aV_k^{Re+})
+V_k^{Re-}\ln(aV_k^{Re-})+\\
+V_k^{Im+}\ln(aV_k^{Im+})+V_k^{Im-}\ln(aV_k^{Im-})+\nonumber\\
+\sum_{n=1}^{n=N}\frac{(\eta_n^{Re})^2+(\eta_n^{Im})^2}{2\sigma_n^2},\nonumber
\end{eqnarray}}
where the sought--for variables $V_k^{Re}$ and $V_k^{Im}$ are
represented as the difference of the positive and negative parts:
$V_k^{Re}=V_k^{Re+}-V_k^{Re-}$ and $V_k^{Im}=V_k^{Im+}-Y_k^{Im-}$
respectively; in this case, $V_k^{Re+}, V_k^{Re-}, V_k^{Im+},
V_k^{Im-}\ge0$, $a\gg 1$ is the real--valued parameter responsible
for the separation of the positive and negative parts of the
sought-for variables with the required accuracy (in our case, we
adopted $a=10^6)$, $K_1$ and $K_2$ are the a priori known lower
and upper localization boundaries of the sought-for finite
spectrum, $\eta_n^{Re}$ and $\eta_n^{Im}$ are a real and an
imaginary parts respectively of the measurement error of the $n$th
value of the series $V_n^{'}$ that obey a random law with a normal
distribution with a zero mean and dispersion $\sigma_n$.

The constraints (\ref{e-19}) on the unknowns, with accounting
measurement errors, can be rewritten as

\begin{eqnarray}
\label{e-21}
 \sum_{k=K1}^{k=K2}
 ((\bar{V}_k^{Re+}-\bar{V}_k^{Re-})+j(\bar{V}_k^{Im+}-\bar{V}_k^{Im-}))\times\\
 \times\exp(j\frac {2\pi (k-1)(l_n-1)}{K})+\eta_n^{Re}+j\eta_n^{Im} = V_{n}^{'}\nonumber.
\end{eqnarray}

We can see from (\ref{e-20}), that the functional to be minimized
consists of five terms, the first four ones are total entropy of
the sought-for solution, the last one is $\chi^2$ measure of
deviation between data and solution. Thus, the GMEM algorithm
seeks for solutions not only for the spectrum unknowns, but also
for measurement errors ($\eta_n^{Re}+j\eta_n^{Im}$). Therefore we
can expect effective suppression of noise caused not only by
non-uniformity of series but also by measurement errors. The
optimization of functional (\ref{e-20}) under conditions
(\ref{e-21}) can be done numerically using any gradient method. We
used a steepest-descent method.

\section{Data}

Bobylev \& Bajkova (2012)  analyzed  a sample of 44 masers with
known high-precision trigonometric parallaxes, line-of-site
velocities and proper motions. Since then the number of masers
with such measurements has increased considerably. Now data on 58
masers are available in literature. Table~1 lists the input data
on 58 masers associated with the youngest Galactic stellar objects
(protostar objects of different mass, very massive supergiants, or
T~Tau stars). The references to majority of original data can be
found in Bajkova \& Bobylev (2012).

The input data, namely, trigonometric parallaxes and proper
motions, were obtained by several groups using long time
radio-interferometric observations carried out within the
framework of different projects. One of them  --- the Japanese
project VERA (VLBI Exploration of Radio Astrometry) (Honma et al.
2007) --- is dedicated to observation of H$_2$O and SiO masers at
22 and 43~GHz, respectively. Note that higher observing frequency
results in higher resolution and more accurate data. Methanol
(CH$_3$OH) masers were observed at 12 GHz (VLBA, NRAO) and 8.4-GHz
continuum radio-interferometric observations of radio stars were
carried out with the same aim (Reid et al. 2009).

\begin{table}
\centering
 {\caption{Data on masers}\label{t1}
      \begin{tabular}{lrrrrrrrrrrr}      \hline

Source&$\alpha$&$\delta$&$\pi$&$\mu_\alpha$&$\mu_\delta$&$V_r$\\\hline

 L1287       &  9.2& 63.5&1.1&  -.9& -2.3&-23\\\hline
 IRAS 00420  & 11.2& 55.8& .5& -2.5&  -.8&-46\\\hline
 NGC281-W    & 13.1& 56.6& .4& -2.7& -1.8&-29\\\hline
 S Per       & 35.7& 58.6& .4&  -.5& -1.2&-38\\\hline
 W3-OH       & 36.8& 61.9& .5& -1.2&  -.2&-44\\\hline
 WB89-437    & 40.9& 63.0& .2& -1.3&   .8&-72\\\hline
 NGC 1333-f12& 52.3& 31.3&4.3& 14.& -8.9&  7\\\hline
 Orion KL    & 83.8& -5.4&2.4&  3.3&   .1& 10\\\hline
 Orion KL SiO& 83.8& -5.4&2.4&  9.6& -3.8&  5\\\hline
 S252A       & 92.2& 21.6& .5&   .1& -2.0& 10\\\hline
 IRAS 06058+ & 92.2& 21.6& .6&  1.1& -2.8&  3\\\hline
 IRAS 06061+ & 92.3& 21.8& .5&  -.1& -3.9& -1\\\hline
 S255        & 93.2& 18.0& .6&  -.1&  -.8&  4\\\hline
 S269        & 93.7& 13.8& .2&  -.4&  -.1& 19\\\hline
 VY CMa      &110.7&-25.8& .8& -2.8&  2.6& 18\\\hline
 G232.62+0.9 &113.0&-17.0& .6& -2.2&  2.1& 22\\\hline
 G14.33-0.64 &274.7&-16.8& .9&   .9& -2.5& 22\\\hline
 G23.43-0.20 &278.7& -8.5& .2& -1.9& -4.1& 97\\\hline
 G23.01-0.41 &278.7& -9.0& .2& -1.7& -4.1& 81\\\hline
 G35.20-0.74 &284.6&  1.7& .5&  -.2& -3.6& 27\\\hline
 W48         &285.4&  1.2& .3&  -.7& -3.6& 41\\\hline
 IRAS 19213+ &290.9& 17.5& .3& -2.5& -6.1& 41\\\hline
 W51         &290.9& 14.5& .2& -2.6& -5.1& 58\\\hline
 V645        &295.8& 23.7& .5& -1.7& -5.1& 27\\\hline
 AFGL2789    &325.0& 50.2& .3& -2.2& -3.8&-44\\\hline
 IRAS 22198  &335.4& 63.9&1.3& -3.0&   .1&-17\\\hline
 L1206       &337.2& 64.2&1.3&   .3& -1.4&-12\\\hline
 CepA        &344.1& 62.0&1.4&   .5& -3.7&-10\\\hline
 NGC7538     &348.4& 61.5& .4& -2.5& -2.4&-57\\\hline
 IRAS 16293  &248.1&-24.5&5.6&-21.&-32.4&  4\\\hline
 L1448C      & 51.4& 30.7&4.3& 22.&-23.1&  4\\\hline
 G 5.89-0.39 &270.1&-24.1& .8&   .2&  -.9& 10\\\hline
 ON1         &302.5& 31.5& .4& -3.1& -4.7& 12\\\hline
 ON2         &305.4& 37.6& .3& -2.8& -4.7&  1\\\hline
 G12.89+0.49 &273.0&-17.5& .4&   .2& -1.9& 39\\\hline
 M17         &275.1&-16.2& .5&   .7& -1.4& 23\\\hline
 G192.16-3.84& 89.6& 16.5& .7&   .7& -1.6&  5\\\hline
 G75.30+1.32 &304.1& 37.6& .1& -2.4& -4.5&-57\\\hline
 W75N        &309.7& 42.6& .8& -2.0& -4.2&  9\\\hline
 DR21        &309.8& 42.4& .7& -2.8& -3.8& -3\\\hline
 DR20        &309.3& 41.6& .7& -3.3& -4.8& -3\\\hline
 IRAS 20290  &307.7& 41.0& .7& -2.8& -4.1& -1\\\hline
 AFGL 2591   &307.7& 41.0& .3& -1.2& -4.8& -5\\\hline
 HW9 CepA    &344.1& 62.0&1.4&  -.7& -1.8&-10\\\hline
 IRAS 5168+36& 80.1& 36.6& .5&   .2& -3.1&-15\\\hline
 NML Cyg     &311.6& 40.1& .6& -1.6& -4.6& -1\\\hline
 IRAS20143+36&304.0& 36.7& .4& -3.0& -4.4&  7\\\hline
 PZ Cas      &356.0& 61.8& .4& -3.2& -2.5&-45\\\hline
 IRAS22480+60&342.5& 60.3& .4& -2.6& -1.9&-50\\\hline
 RCW 122     &260.0&-39.0& .3&  -.7& -2.8&-12\\\hline
 Hubble 4    & 64.7& 28.3&7.5&  4.3&-28.9& 15\\\hline
 HDE 283572  & 65.5& 28.3&7.8&  8.9&-26.6& 15\\\hline
 TTau N      & 65.5& 19.5&6.8& 12.&-12.8& 19\\\hline
 V773 Tau AB & 63.6& 28.2&7.7&  8.3&-23.6& 16\\\hline
 HP TG2      & 69.9& 22.9&6.2& 14.&-15.4& 17\\\hline
 S1     Oph  &246.6&-24.4&8.6& -3.9&-31.5&  3\\\hline
 DoAr21 Oph  &246.5&-24.4&8.2&-26.&-28.2&  3\\\hline
 EC 95       &277.5&  1.2&2.4&   .7& -3.6&  9\\\hline
\end{tabular}}
\end{table}

The line-of-site velocities $V_r(LSR)$ listed in Table 1 were
determined with respect to the Local Standard of Rest by different
authors from radio observations in CO emission lines.  The
parallaxes were determined, on average,  with a relative error of
$\sigma_\pi/\pi\approx5\%,$ and only in three regions the error
exceeds the mean level. These are IRAS~16293-2422
($\sigma_\pi/\pi=19\%$), G~23.43-0.20 ($\sigma_\pi/\pi=18\%$) and
W~48 ($\sigma_\pi/\pi=14\%$).

\begin{figure}
\includegraphics[width=70mm]{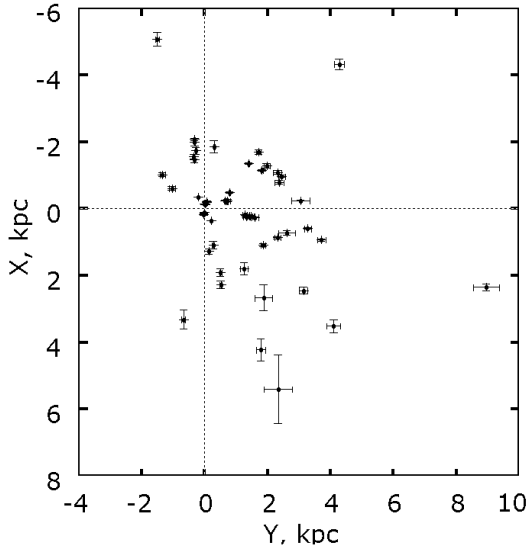}
\caption{Coordinates of masers in the $XY$ Galactic plane (the Sun
is located at the center of coordinate system).}
\label{f-0}
\end{figure}

Fig.\, \ref{f-0} shows the space distribution of masers projected
onto the Galactic $XY$ plane. The objects can be seen to be widely
scattered along the $x$, $y$ coordinates, implying a large scatter
of position angles. Hence the spectrum analysis algorithm has to
be constructed in order to correctly extract periodic signal from
velocity perturbations traced by masers.

\begin{figure}
\includegraphics[width=70mm]{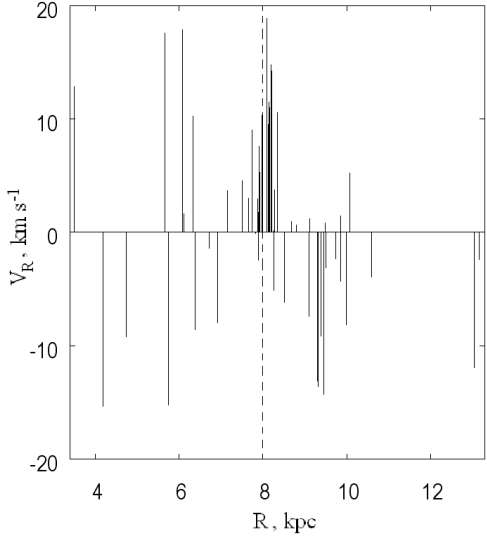}
\caption{Masers galactocentric radial velocities vs galactocentric
distances $R$.} \label{f-2}
\end{figure}

\section{Results}

First, we re-determined the parameters of the Galactic rotation
curve using the data for 58 masers. The method employed is based
on the well-known Bottlinger formulae (Ogorodnikov, 1965), where
the angular velocity of Galactic rotation is expanded into a
series up to 2-nd order terms in $r/R_0$ (Bobylev \& Bajkova
2010). We adopt $R_0=8$~kpc and infer the following components of
the peculiar solar velocity:
 $(U_\odot,V_\odot,W_\odot)=(7.4,16.6,8.53)\pm(1.0,0.8,0.5)$~km c$^{-1}$,
and the following parameters of the Galactic rotation curve:
 $\Omega_0=-29.3 \pm0.6,$~km c$^{-1}$ kpc $^{-1}$,
 $\Omega'_0=+4.2\pm0.1$~km c$^{-1}$ kpc $^{-2}$,
 $\Omega''_0=-0.85\pm0.03,$~km c$^{-1}$ kpc $^{-3}$.
The linear Galactic rotation velocity at $R=R_{\circ}$ then is
equal to: $V_0=|R_0\Omega_0|=234\pm5$~km c$^{-1}$.

There is good agreement of our results with the results of
analyzing masers by different authors. Based on a sample of 18
masers, McMillan \& Binney (2010) showed that $\Omega_0$ lying
within the range $29.9-31.6$~km s$^{-1}$ kpc$^{-1}$ at various
$R_0$ was determined most reliably and obtained an estimate of
$V_0=247\pm19$~km s$^{-1}$ for $R_0=7.8\pm0.4$~kpc. Based on a
sample of 18 masers, Bovy, Hogg \& Rix  (2009) found
$V_0=244\pm13$~km s$^{-1}$ at $R_0=8.2$~kpc. Using 44 masers,
Bajkova \& Bobylev (2012) found:
 $(U_\odot,V_\odot,W_\odot)=(7.6,17.8,8.3)\pm(1.5,1.4,1.2)$~km s$^{-1}$,
 $\Omega_0=-28.8 \pm0.8$~km s$^{-1}$ kpc$^{-1}$,
 $\Omega'_0=+4.18\pm0.15$~km s$^{-1}$ kpc$^{-2}$,
 $\Omega''_0=-0.87\pm0.06$~km s$^{-1}$ kpc$^{-3}$,
$V_0=|R_0\Omega_0|=230\pm14$~km s$^{-1}$.

It is important that the rotation-curve parameters found are in
good agreement with the results of analyzing young Galactic disk
objects rotating most rapidly around the center: OB associations
with $\Omega_0 =-31\pm1$~km s$^{-1}$ kpc$^{-1}$ (Mel'nik, Dambis,
\& Rastorguev 2001; Mel'nik \& Dambis 2009), blue supergiants with
$\Omega_0=-29.6\pm1.6$~km s$^{-1}$ kpc$^{-1}$ and
   $\Omega'_0= 4.76\pm0.32$~km s$^{-1}$ kpc$^{-2}$ (Zabolotskikh, Rastorguev \& Dambis 2002),
   or~OB3 stars with
 $\Omega_0 = -31.5\pm0.9$~km s$^{-1}$ kpc$^{-1}$,
 $\Omega^{'}_0 = +4.49\pm0.12$~km s$^{-1}$ kpc$^{-2}$ and
 $\Omega^{''}_0 = -1.05\pm0.38$~km s$^{-1}$ kpc$^{-3}$ (Bobylev \& Bajkova 2011).

The galactocentric radial, $V_{R_n},$ and tangential,
$V_{\theta_n}$ ($n=1,\dots,44,)$ velocities of the masers were
determined from the relations
 \begin{equation}
 V_{\theta_n}= U_n\sin \theta_n+(V_0+V_n)\cos \theta_n,\label{e-22}
 \end{equation}
\begin{equation}
  V_{R_n}=-U_n\cos \theta_n+(V_0+V_n)\sin \theta_n,\label{e-23}
\end{equation}
where  $U_n, V_n$ are the heliocentric space velocities adjusted
for solar peculiar motion.

The residual tangential velocities $\Delta V_{\theta_n}$ are
obtained from the tangential velocities (\ref{e-22}) minus the
smooth rotation curve that is defined by the Galactic rotation
parameters $\Omega_0,$ $\Omega'_0,$ and $\Omega''_0$ found. The
radial velocities (\ref{e-23}) depend only on one Galactic
parameter $\Omega_0$ and do not depend on the rotation curve. As
our experience showed, the data are so far insufficient to
reliably extract the density wave from the tangential residual
velocities of the masers. Therefore, here we determine the spiral
density wave parameters only from galactocentric radial
velocities.

\begin{figure}
\includegraphics[width=70mm]{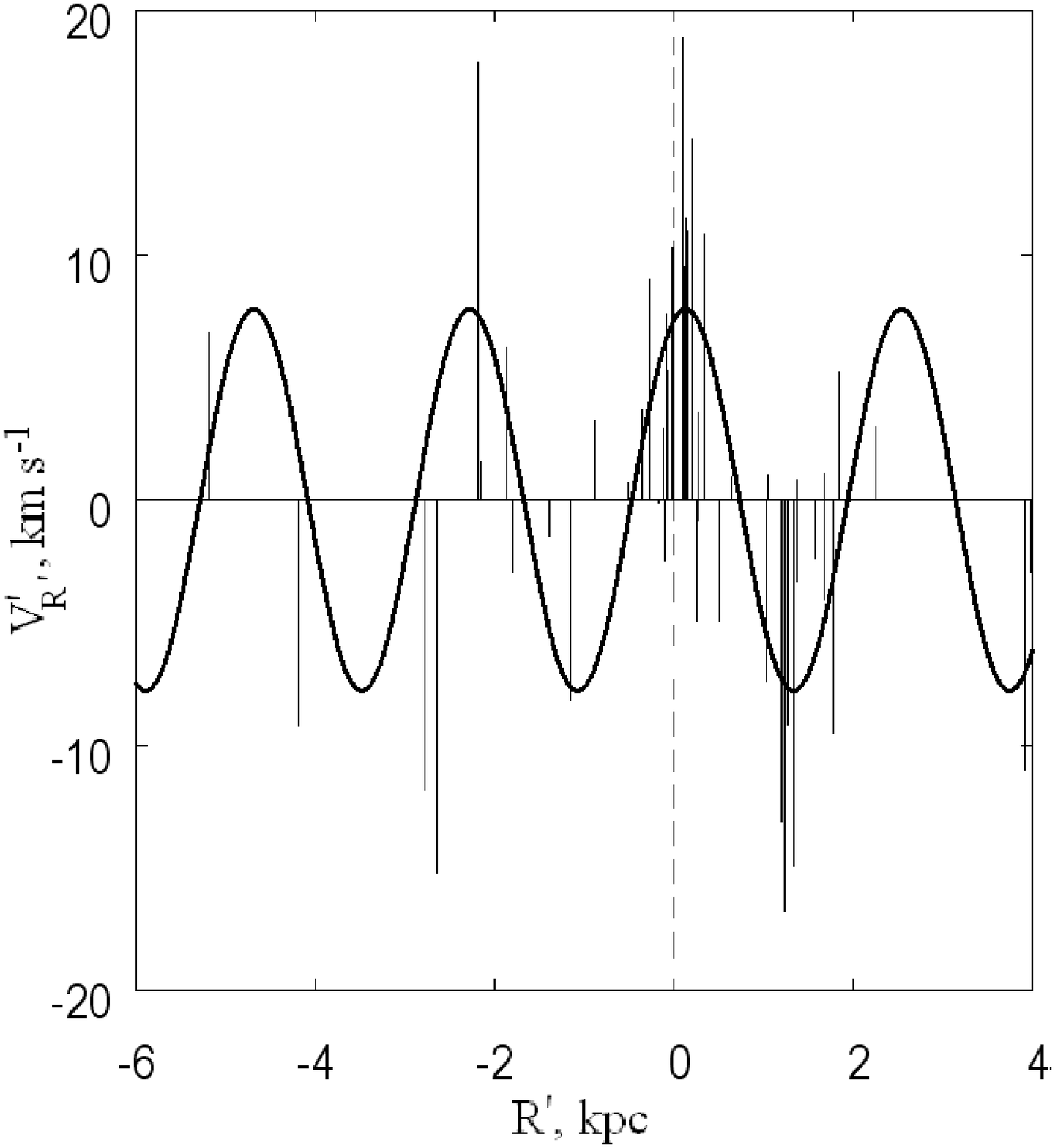}
\caption{Transformed galactocentric radial velocities vs $R^{'}$
and extracted main harmonic fitted to the data (solid bold line).}
\label{f-3}
\end{figure}

\begin{figure}
\includegraphics[width=70mm]{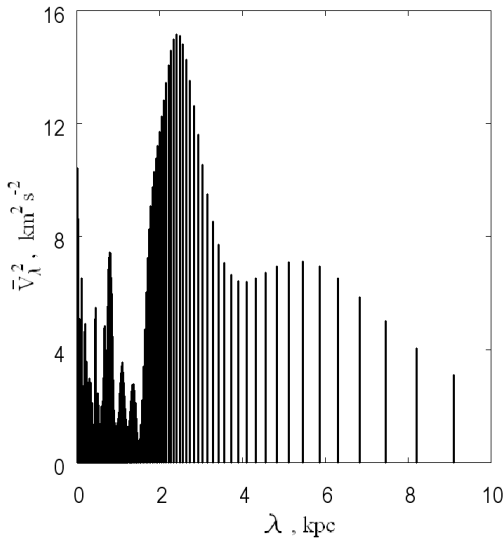}
\caption{Periodogram (spectrum power) of masers galactocentric
radial velocities.} \label{f-4}
\end{figure}

Figure~\ref{f-2} shows the input radial velocity series $V_{R_n},
n=1,\dots,58$. Figure~\ref{f-3} shows transformed velocity series
$V^{'}_{R^{'}_n}, n=1,\dots,58$ and main extracted harmonic. The
periodogram obtained using a modified Fourier transform-based
method is given in fig.~\ref{f-4}. An analysis of this periodogram
allowed us to estimate the following spiral wave parameters: the
amplitude of the radial perturbations $f_R=7.5 \pm 1.5$ km
c$^{-1}$; the wavelength (interarm distance in the galactocentric
direction) $\lambda=2.4 \pm 0.4$ kpc, and the phase of the Sun in
the spiral density wave $\chi_\odot=-160 \pm 15^\circ$. The pitch
angle of the spiral wave estimated from equation (\ref{e-04}) for
$m=2$ is $-5.5 \pm 1^\circ$. The significance level of the peak is
$p=0.99$. Significance level was estimated using the simplest
method based on Schuster theorem (Vityazev, 2001).The error bars
are based on a Monte-Carlo simulation of 1000 random realizations
of input data assuming that measurement errors obey the normal
distribution. Note that the value for $f_R$ found by fitting the
harmonic with $\lambda=2.4$ kpc to data is in good agreement with
relation (\ref{e-18}).

\begin{figure}
\includegraphics[width=70mm]{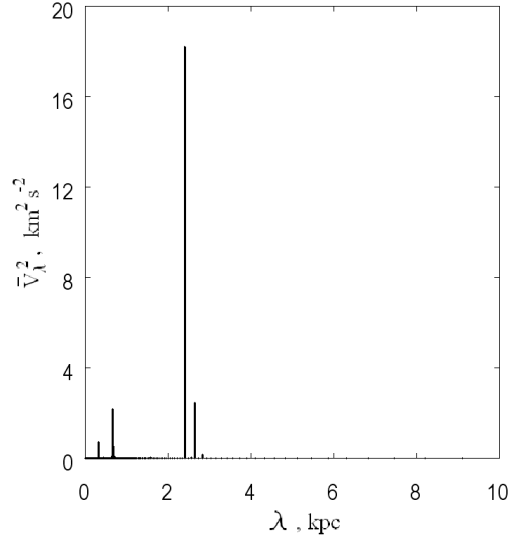}
\caption{The spectrum power reconstructed by the GMEM.}
\label{f-6}
\end{figure}

Power spectrum, obtained using the GMEM, is shown in
fig.\,\ref{f-6}. As we can see, this nonlinear reconstruction
algorithm allowed us to get rid of the side lobes near the main
peak almost completely and, thus, to increase considerably the
significance of the extracted periodicity $(p=1.0)$ with
$\lambda=2.4$ kpc.

For comparison, in our previous study (Bajkova \& Bobylev 2012) we
have obtained from data on 44 masers the following spiral density
wave parameters: amplitude $f_R = 7.7\pm 1.7$ km s$^{-1}$,
wavelength $\lambda=2.2^\pm 0.4$~kpc, pitch angle $i=-5\pm
0.9^\circ$, and the phase of the Sun $\chi_\odot= -147\pm
17^\circ$.

The parameters of Galactic spiral density wave obtained are in
good agreement with those found by different authors by applying
different methods to different Galactic tracer objects (Mel'nik et
al.~2001; Zabolotskikh et al.~2002; Bobylev \& Bajkova 2011 and
many others).

\section{Conclusions}

We used both Fourier transform-based spectral analysis technique
and the generalized maximum entropy reconstruction method to
extract a periodic signal from the galactocentric radial
velocities of 58 masers with currently known high-precision
trigonometric parallaxes, proper motions and line-of-site
velocities. In accordance with Lin \& Shu (1964) theory, the
extracted periodic signal is associated with the Galactic spiral
density wave. Masers span a wide range of galactocentric distances
$3<$R$<14$~kpc and show a large scatter of position angles
$\theta$ in the Galactic $XY$ plane, making it necessary an exact
accounting for both the logarithmic nature of the argument and
position angles. As a result, we re-determined the main parameters
of the Galactic spiral density wave as follows: amplitude of
radial velocities perturbations $f_R=7.5 \pm 1.5$ km s$^{-1}$,
wavelength $\lambda=2.4 \pm 0.4$ kpc, pitch angle $-5.5 \pm
1^\circ$ and phase of the Sun in the density wave $\chi_\odot=-160
\pm 15^\circ$.

\acknowledgements This work was supported by the ``Non -
stationary processes in the Universe'' Program of the Presidium of
the Russian Academy of Science and the Program of State Support
for Leading Scientific Schools of the Russian Federation (project.
NSh--16245.2012.2, ``Multi-wavelength Astrophysical Studies'').

\end{document}